\documentclass[twocolumn,showpacs,prl,amsmath,amssymb,floatfix,10pt,aps]{revtex4-1}
\usepackage{amsmath,amssymb,color}
\usepackage{bm}
\usepackage{graphicx}
\usepackage{psfrag}
\usepackage{hyperref}
\usepackage[all]{hypcap}
\usepackage{booktabs,array}
\newcolumntype{C}{>{$}c<{$}}
\AtBeginDocument{
	\heavyrulewidth=.08em
	\lightrulewidth=.05em
	\cmidrulewidth=.03em
	\belowrulesep=.65ex
	\belowbottomsep=0pt
	\aboverulesep=.4ex
	\abovetopsep=0pt
	\cmidrulesep=\doublerulesep
	\cmidrulekern=.5em
	\defaultaddspace=.5em
}
\newcommand{\ra}[1]{\renewcommand{\arraystretch}{#1}}
\newcommand{\bd}{\bm}

\begin{document}

\title{Exact renormalization group for quantum spin systems}

\author{Jan Krieg and Peter Kopietz}
  
\affiliation{Institut f\"{u}r Theoretische Physik, Universit\"{a}t
  Frankfurt,  Max-von-Laue Stra{\ss}e 1, 60438 Frankfurt, Germany}

%\date{\today}
\date{December 19, 2018}

 \begin{abstract}

We show that the diagrammatic approach to quantum spin systems developed in a seminal work by Vaks, Larkin, and Pikin [Sov. Phys. JETP {\bf{26}}, 188 (1968)] can be embedded in the framework of the functional renormalization group. The crucial insight is that the generating functional of the time-ordered connected spin correlation functions of an  arbitrary quantum spin system satisfies an exact renormalization group flow equation which resembles the corresponding flow equation of interacting bosons. The $SU (2)$ spin algebra is implemented via a non-trivial initial condition for the renormalization group flow. Our method is rather  general and offers a new non-perturbative approach to quantum spin systems.

\end{abstract}

\maketitle

{\it{Introduction.}} Quantum spin models play a central role in condensed matter physics 
and statistical mechanics
for gaining a  microscopic understanding of the magnetic properties of insulators
with localized magnetic moments~\cite{Auerbach94,Schollwoeck04}. Although theoretical research in this field has a long history starting
with  the seminal papers by Ising \cite{Ising25} and Bethe \cite{Bethe31}, 
the controlled calculation of the physical properties of realistic quantum spin models
describing experimentally accessible materials
remains a highly relevant problem of general interest.  
This is especially challenging in reduced dimensions, where the effect of fluctuations can be sufficiently strong to destroy any  long-range magnetic order.
But also  in three dimensions, competing interactions or geometrical frustration
can destroy long-range magnetic order and stabilize exotic states
characterized by topological order \cite{Balents10,Castelnovo08}.

The low-energy excitations of ordered magnets are usually renormalized spin-waves. 
In this case
an  expansion in powers of the inverse spin-quantum number $1/S$,
formalized
with the help of the Holstein-Primakoff~\cite{Holstein40}  or the Dyson-Maleev \cite{Dyson56,*Maleev58} transformation, 
has been extremely successful and continues to be one of the 
most powerful theoretical methods for ordered 
magnets \cite{Zhitomirsky13}.
However, in the absence of  long-range magnetic order the $1/S$-expansion
is not applicable. Several alternative  methods have been developed
to study  quantum magnets without magnetic order, such as modifications of
spin-wave theory where the vanishing magnetization is externally enforced \cite{Takahashi86,*Takahashi87a,*Takahashi87b,Kollar03}, 
Schwinger-boson mean-field theory \cite{Arovas88,Auerbach94}, and 
mean-field theories relying on the
representation of the spin operators in terms of 
Abrikosov pseudofermions \cite{Abrikosov65,Coleman15} or
Majorana fermions \cite{Tsvelik92,Shnirman03,Biswas11,Herfurth13}.

Each of the above methods has its own shortcomings.
While the Majorana representation of spin operators 
generates redundancy in Hilbert space \cite{Biswas11}, the pseudofermion representation as well as the Schwinger-boson approach introduce
unphysical states which should be projected out.
In practice, this projection can only be implemented
approximately. For pseudofermions this can be achieved
using a method due to Popov and Fedotov \cite{Popov88}, who showed
that the contribution from  unphysical states 
cancels if one introduces a certain
imaginary chemical potential \cite{Kiselev00}. 
Recently Reuther and W\"{o}lfle \cite{Reuther10}  have developed a
functional renormalization group (FRG) \cite{Berges02,Kopietz10,Metzner12}
approach for spin-$1/2$  systems using the pseudofermion representation, which has been quite successful to understand the phase diagram of various frustrated magnets \cite{Reuther11A,Reuther11B,Reuther11C,Singh12}.

In this work, we shall develop an alternative FRG approach 
for quantum spin models with arbitrary spin $S$ 
which does not rely on any auxiliary representation of the
spin operators.
The main idea is to  formulate the FRG directly in terms of 
the physical spin operators, thus avoiding the introduction of
fermionic or bosonic auxiliary operators acting on a projected Hilbert space.
In the recent work~[\onlinecite{Werth18}] this strategy has been adopted
to study  low-dimensional 
$S=1/2$ quantum antiferromagnets within a mean-field decoupling.
In fact, an approach to quantum spin systems 
which works directly with the physical spin operators
has been developed half a century ago by Vaks, Larkin, and Pikin (VLP) \cite{Vaks68a,*Vaks68b}, who 
showed that the  
spin operators satisfy a generalized Wick theorem, which can be used to 
develop a systematic diagrammatic expansion in powers of the inverse range of the exchange interaction. 
A  detailed description of  the VLP approach can be found in a textbook by  Izyumov and
Skryabin \cite{Izyumov88}.
Although this method has been further developed \cite{Maleev74,Izyumov02}, 
it has not gained a wide popularity, perhaps because of the rather cumbersome
diagrammatic rules implied by the generalized Wick theorem for spin operators.
In this work we show that by embedding the VLP idea into the framework of the
FRG, we can avoid this technical problem and
obtain a powerful analytical approach
to quantum spin systems.

{\it{Exact flow equations.}}
Although our method can easily be extended to more general spin models, we consider here for simplicity the
quantum Heisenberg Hamiltonian
 \begin{equation}
 {\cal{H}} = 
 \frac{1}{2} \sum_{ij} J_{ij} {\bd{S}}_i \cdot {\bd{S}}_j  - h_0  
 \sum_i S^z_i  ,
 \label{eq:hamiltonian}
 \end{equation}
where the subscripts $i,j$ label the $N$ sites $\bd{r}_i$ of a 
$D$-dimensional lattice,
$h_0$ is an external magnetic field in units of energy,
and $J_{ij} = J ( \bd{r}_i - \bd{r}_j )$ are arbitrary exchange couplings. 
The spin-$S$ operators $\bd{S}_i$ are normalized such that $\bd{S}_i^2 = S (S+1)$ 
and satisfy the usual $SU (2)$-algebra
$ [ S_i^{\alpha} , S_j^{\beta} ] = i \delta_{ i j} \epsilon^{\alpha \beta \gamma} S_i^{\gamma}$,
where the superscripts $\alpha, \beta, \gamma$ refer to the Cartesian components of
$\bd{S}_i$
and $\epsilon^{\alpha \beta \gamma}$ is the
totally antisymmetric $\epsilon$-tensor.
We now replace the exchange couplings $J_{ij}$ by some 
continuous deformation
$J^{\Lambda}_{ij}$, which depends on a 
dimensionless parameter $ \Lambda \in [ 0,1]$ such that
$J^{\Lambda =0}_{ij}$ is sufficiently simple to allow for a
controlled solution of the initially deformed spin model, and
$J^{\Lambda =1}_{ij} = J_{ij}$ so that for $\Lambda =1$ we recover our original model.

To begin with, we derive an exact evolution equation for the $\Lambda$-dependent
generating functional
of the connected  Euclidean time-ordered spin correlation functions,
  \begin{equation}
 {\cal{G}}_{\Lambda} [ \bd{h} ]
= \ln {\rm Tr} \left[ e^{ - \beta {\cal{H}}_0 } {\cal{T}} e^{ \int_0^{\beta} 
 d \tau    [    \sum_i \bd{h}_i ( \tau ) \cdot {  {\bd{S}}_i ( \tau )  -  
 {\cal{V}}_{\Lambda} ( \tau )   ] } }  \right].
 \label{eq:Gcdef}
 \end{equation}
Here $\beta$ is the inverse temperature, ${\cal{T}}$ denotes time-ordering in imaginary time,
$\bd{h}_i ( \tau )$ are fluctuating source fields, ${\cal{H}}_0 = - h_0 \sum_i S^z_i$ is the local
part of the spin Hamiltonian,
${\cal{V}}_{\Lambda} ( \tau ) = \frac{1}{2} \sum_{ij} J^{\Lambda}_{ij} {\bd{S}}_i ( \tau ) \cdot {\bd{S}}_j ( \tau )$ is the deformed exchange Hamiltonian, and the time dependence of all operators is in the interaction picture with respect to ${\cal{H}}_0$. 
The connected time-ordered spin correlation functions can be obtained by taking 
derivatives of  ${\cal{G}}_{\Lambda} [ \bd{h} ]$ with respect to the sources.
For example, the local magnetic moment  at 
lattice site $\bd{r}_i$ is given by
 $\langle \bd{S}_i ( \tau ) \rangle_\Lambda  =   \left.  \delta {\cal{G}}_{\Lambda} [ \bd{h} ]  / \delta \bd{h}_i ( \tau ) 
 \right|_{ \bd{h} =0 }$,
and the connected time-ordered spin-spin correlation function can be generated as follows,
 \begin{eqnarray}
 G_{\Lambda,ij}^{\alpha \alpha^{\prime}} ( \tau , \tau^{\prime} ) & = &
 \langle {\cal{T}} \bigl[ S_i^{\alpha} ( \tau ) 
 S_j^{\alpha^{\prime}} ( \tau^{\prime} ) \bigr]
\rangle_\Lambda    -  \langle S_i^{\alpha} ( \tau ) \rangle_\Lambda
 \langle S_j^{\alpha^{\prime}} ( \tau^{\prime} ) \rangle_\Lambda
 \nonumber
 \\
 & = & 
 \left. \frac{ \delta^2 {\cal{G}}_{\Lambda} [ \bd{h} ] }{\delta h_i^{\alpha} ( \tau )   \delta h_j^{\alpha^{\prime}} ( \tau^{\prime} ) } 
 \right|_{\bd{h} =0} .
 \label{eq:G2def}
 \end{eqnarray}
By simply differentiating Eq.~(\ref{eq:Gcdef}) with respect to the deformation parameter $\Lambda$ we 
obtain the  exact flow equation 
 \begin{eqnarray}
\partial_{\Lambda}{\cal{G}}_{\Lambda} [ \bd{h} ]  
 &  = & -
\frac{1}{2} \int_0^{\beta} d \tau 
 \sum_{ij, \alpha} ( \partial_{\Lambda} J^{\Lambda}_{ij} ) 
 \Biggl[\frac{ \delta^2 {\cal{G}}_{\Lambda} [ \bd{h} ] }{\delta h_i^{\alpha} ( \tau )   
\delta h_j^{\alpha} ( \tau ) }
 \nonumber
 \\
 & & \hspace{16mm}
 + 
\frac{ \delta {\cal{G}}_{\Lambda} [ \bd{h} ] }{\delta h_i^{\alpha} ( \tau ) }
\frac{ \delta {\cal{G}}_{\Lambda} [ \bd{h} ] }{\delta h_j^{\alpha} ( \tau ) }
  \Biggr].
 \label{eq:flowW}
 \end{eqnarray}
Note that in the derivation of FRG  flow equations for
interacting field theories, it is usually assumed that the relevant generating functional  
can be represented 
in terms of some unconstrained functional integral over real, complex,  
or Grassmann fields \cite{Berges02,Kopietz10,Wetterich93}. However, this assumption is really not 
necessary, as pointed out before by Pawlowski \cite{Pawlowski07}, see also
Refs.~[\onlinecite{Machado10,Rancon14,Krieg17}]. This insight is crucial for applying FRG techniques to 
models defined in terms of operators satisfying  neither bosonic nor 
fermionic commutation relations.
The exact flow equation (\ref{eq:flowW}) is equivalent to an infinite hierarchy of flow equations for 
the connected time-ordered $n$-spin correlation functions $G^{\alpha_1 \ldots \alpha_n}_{\Lambda, 
 i_1 
\ldots i_n} ( \tau_1 , \ldots , \tau_n )$, which are defined via the derivatives 
of  ${\cal{G}}_{\Lambda} [ \bd{h} ]  $ with respect to the sources ${h}^{\alpha}_i ( \tau )$.
The hierarchy of flow equations can be written as
 \begin{widetext}
 \begin{eqnarray}
  & & \partial_{\Lambda}    G_{\Lambda, i_1 \ldots i_n }^{\alpha_1 \ldots \alpha_n } 
 ( \tau_1, \ldots,  \tau_n )   =   -  \frac{1}{2} \int_0^{\beta} d \tau 
 \sum_{ij, \alpha} ( \partial_{\Lambda} J^{\Lambda}_{ij} )
 \biggl[G_{\Lambda, i_1 \ldots i_n ij }^{\alpha_1 \ldots \alpha_n \alpha \alpha  } 
 ( \tau_1, \ldots,  \tau_n, \tau , \tau  )
 \nonumber
 \\
 & &  + \sum_{ m=0}^n  {\cal{S}}_{ 1, \ldots, m; m+1, \ldots, n }
 \left\{
G_{\Lambda, i_1 \ldots i_m i }^{\alpha_1 \ldots \alpha_m \alpha  } 
 ( \tau_1, \ldots,  \tau_m, \tau   )
G_{\Lambda, i_{m+1} \ldots i_n j }^{\alpha_{m+1} \ldots \alpha_n \alpha  } 
 ( \tau_{m+1}, \ldots,  \tau_n, \tau   )
 \right\} \biggr],
 \label{eq:flowGn}
 \end{eqnarray}
 \end{widetext}
where the symmetrization operator ${\cal{S}}_{ 1, \ldots, m; m+1, \ldots, n } \left\{ \ldots \right\}$ 
symmetrizes the expression in the curly braces with respect to the
exchange of all labels \cite{Kopietz10}.
 A graphical representation of Eq.~(\ref{eq:flowGn})
is shown in Fig.~\ref{fig:flown}.
\begin{figure}[tb]
 \begin{center}
  \centering
\vspace{7mm}
 \includegraphics[width=\linewidth]{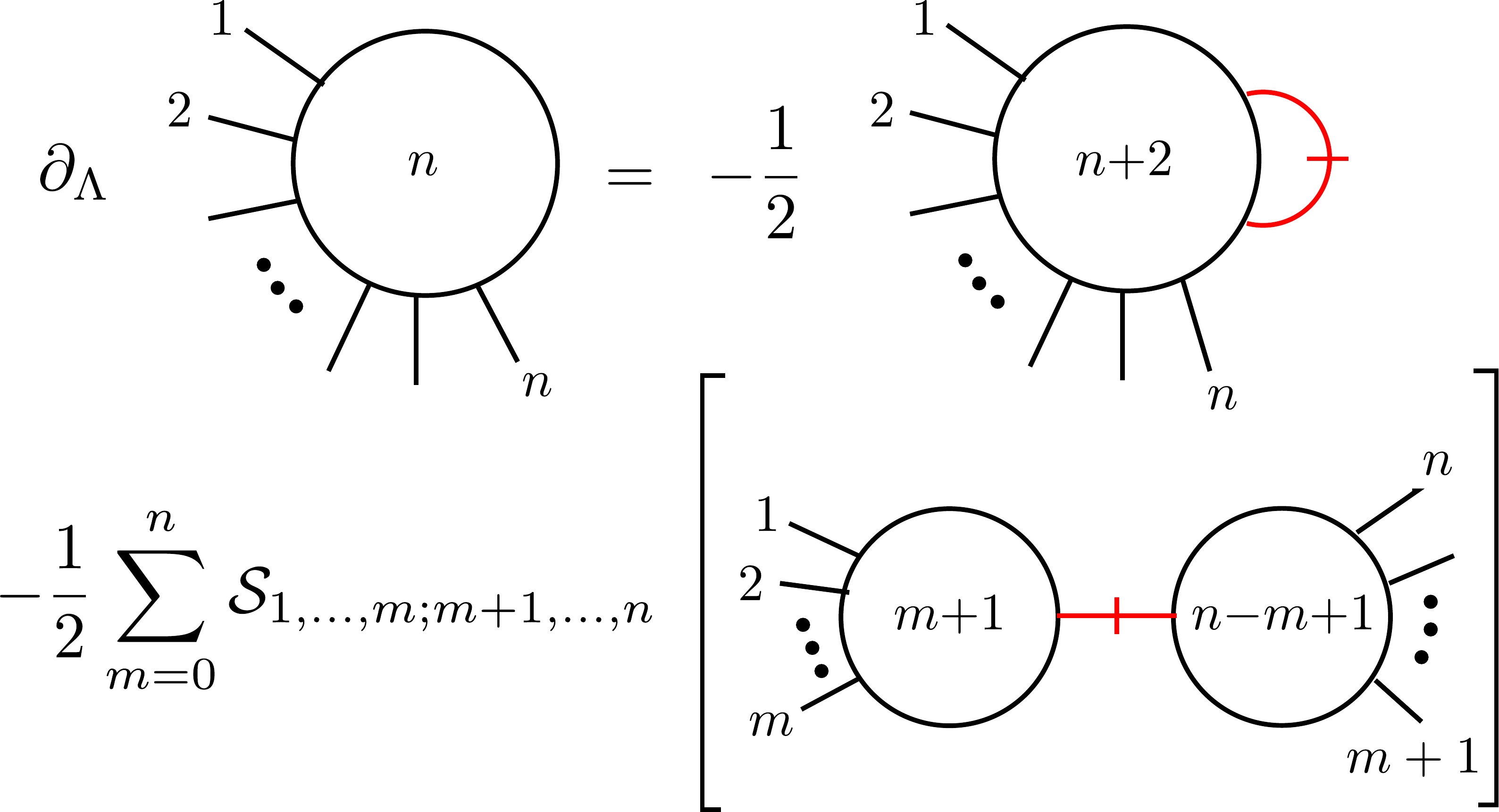}
   \end{center}
  \caption{
Graphical representation of the exact flow equation (\ref{eq:flowGn})
 for the connected time-ordered $n$-spin correlation functions, which 
are represented by  circles with $n$ external legs. 
Here the red, slashed lines denote the derivative
$ \partial_{\Lambda} J_{ij}^{\Lambda}$ of the deformed exchange coupling.
}
\label{fig:flown}
\end{figure}
The exact flow equation (\ref{eq:flowGn}) can be used to generate a systematic expansion
of the connected spin correlation functions in powers of the exchange couplings. 
Therefore we choose the deformation scheme $J^{\Lambda}_{ i j } = \Lambda J_{ij}$, so that
each slashed line in Fig.~\ref{fig:flown} gives simply an additional power of $J_{ij}$.
A straightforward iteration of the system of flow equations then generates the desired expansion.
This algorithm seems to be considerably simpler than the method 
based on the generalized Wick theorem for spin operators \cite{Izyumov02}.

Following the usual procedure \cite{Wetterich93,Berges02,Kopietz10,Metzner12}, we now introduce the
generating functional $\Gamma_{\Lambda} [ \bd{M} ]$ of the irreducible spin vertices via
a subtracted Legendre transformation of ${\cal{G}}_{\Lambda} [ \bd{h} ]$,
\begin{eqnarray}
 \Gamma_{\Lambda} [ \bd{M} ] & = &
  \int_0^{\beta} d \tau \sum_i \bd{h}_i ( \tau ) \cdot \bd{M}_i ( \tau )
  - {\cal{G}}_{\Lambda} [ \bd{h}  ]
 \nonumber
 \\
 &- &  \frac{1}{2} \int_0^{\beta} d \tau \sum_{ij} R^{\Lambda}_{ij}
 \bd{M}_i ( \tau ) \cdot \bd{M}_j ( \tau ),
 \label{eq:Gammadef}
 \end{eqnarray}
where $R_{ij}^{\Lambda} = J^{\Lambda}_{ij} - J_{ij}$ 
plays the role of a regulator function \cite{Berges02,Kopietz10,Metzner12}.
Taking a  derivative of $\Gamma_{\Lambda} [ \bd{M} ]$ 
with respect to $\Lambda$ and using
Eq.~(\ref{eq:flowW}) we obtain
 \begin{equation}
  \partial_{\Lambda} \Gamma_{\Lambda} [ \bd{M} ]
  =  \frac{1}{2} \int_0^{\beta} d \tau \sum_{ij, \alpha} ( \partial_{\Lambda} R^{\Lambda}_{ij} )
  \frac{ \delta^2 {\cal{G}}_{\Lambda} [ \bd{h} ] }{\delta h_i^{\alpha} ( \tau )   
 \delta h_j^{\alpha} ( \tau ) }.
 \label{eq:flowGW}
 \end{equation}
By construction, the last term in Eq.~(\ref{eq:flowGW})  can be expressed via
the second functional derivative of  $\Gamma_{\Lambda} [ \bd{M} ]$,
 \begin{eqnarray}
 \frac{ \delta^2 {\cal{G}}_{\Lambda} [ \bd{h} ] }{\delta h_i^{\alpha} ( \tau )   
 \delta h_j^{\alpha} ( \tau ) } 
 & = &   \left[ \mathbf{\Gamma}^{\prime \prime}_{\Lambda} [ \bd{M} ]  + \mathbf{R}_{\Lambda} \right]^{-1}_{i \tau \alpha,
j  \tau  \alpha },
 \end{eqnarray}
where $\mathbf{\Gamma}^{ \prime \prime}_{\Lambda}$ and $\mathbf{R}_{\Lambda}$ are matrices in all
labels with matrix elements
 \begin{eqnarray}
 {\bigl[} 
 \mathbf{\Gamma}^{ \prime \prime }_{\Lambda} [ \bd{M} ]  {\bigr]}_{ i \tau \alpha, j \tau^{\prime} \alpha^{\prime} } 
 & = & 
  \frac{ \delta^2 \Gamma_{\Lambda} [ \bd{M} ] }{\delta M_i^{\alpha} ( \tau ) 
\delta M_j^{\alpha^{\prime}} ( \tau^{\prime} )  },
 \\
 {\bigl[} 
 \mathbf{R}_{\Lambda} {\bigr]}_{ i \tau \alpha, j \tau^{\prime} \alpha^{\prime} } 
 & = & R_{ij}^{\Lambda} \delta_{ \alpha \alpha^{\prime}} \delta ( \tau - \tau^{\prime} ).
 \end{eqnarray}
With this notation the flow equation  (\ref{eq:flowGW})
can be written in the compact form \cite{Wetterich93,Berges02,Kopietz10}
 \begin{equation}
 \partial_{\Lambda} \Gamma_{\Lambda} [ \bd{M} ] = \frac{1}{2} {\rm Tr} \left\{
 \left(
 \mathbf{\Gamma}^{\prime \prime }_{\Lambda} [ \bd{M} ]  + \mathbf{R}_{\Lambda} \right)^{-1}
   \partial_{\Lambda} \mathbf{R}_{\Lambda} 
 \right\}.
 \label{eq:Wetterich}
 \end{equation}
We thus arrive at the important conclusion that, 
in spite of the fact that
the time-ordered spin correlation functions cannot be
represented in terms of an unconstrained functional integral 
over bosonic or fermionic fields,
the  generating functional of the irreducible spin vertices 
satisfies an exact  flow equation which is formally identical to
the bosonic version of  the Wetterich equation~\cite{Wetterich93}.
The bosonic nature of time-ordered spin correlation functions is a direct consequence of the fact that
they satisfy bosonic  Kubo-Martin-Schwinger
boundary conditions \cite{Izyumov88}.
The algebraic structure encoded in the Wetterich equation
therefore also describes the FRG flow of the irreducible vertices 
of  quantum spin systems.
A similar simplification does not occur in the spin-diagram 
technique \cite{Vaks68a,*Vaks68b,Izyumov88,Izyumov02},
where the generalized Wick theorem for spin operators
leads to rather complicated diagrammatic rules
which do not resemble the usual rules for bosons or fermions.
In our spin functional renormalization group (SFRG) approach,
the $SU(2)$ spin algebra is fully taken into account  
via a  non-trivial initial condition 
involving infinitely many higher-order vertices.

{\it{Ising limit.}}  
To understand the origin of the non-trivial initial condition in our
SFRG, it is instructive to consider first the
Ising limit where all operators commute, so that the time-ordering symbol
in Eq.~(\ref{eq:Gcdef}) can be omitted and all
spin correlation functions and irreducible vertices are independent of time.
For special deformation schemes where initially $J^{\Lambda=0}_{ij} = 0$, our SFRG reduces then to the lattice FRG scheme for classical spin models developed by Machado and Dupuis \cite{Machado10}.
For our purpose, it is sufficient to work with the deformed interaction
$J^{\Lambda}_{ij} = \Lambda J_{ij}$.
The Hamiltonian of the
spin-$S$ Ising model with ferromagnetic nearest-neighbor coupling $J$  can be obtained by 
replacing the operator
${\cal{V}}_{\Lambda} ( \tau )$ in Eq.~(\ref{eq:Gcdef}) by
 ${\cal{V}}_{\Lambda} = - \Lambda J \sum_{\langle ij \rangle } S^z_i S^z_j $,
 where
$\langle ij \rangle$ denotes all distinct pairs of nearest neighbors
on a $D$-dimensional hypercubic lattice. The magnetic field $\bd{h}_i $ and the conjugate 
magnetization
$\bd{M}_i$ have then only $z$-components, which we denote by $h_i$ and $M_i$. 
In momentum space
the vertex expansion of $\Gamma_{\Lambda} [ M ]$ is
\begin{eqnarray}
 {\Gamma}_{\Lambda} [ M]  
 & =  &
  \sum_{n=0}^{\infty} \frac{1}{n! N^{n-1}} \sum_{ \bd{k}_1 \ldots \bd{k}_n }
 \delta_{ \bd{k}_1 + \ldots + \bd{k}_n, 0} 
 \nonumber
 \\
 & & \times
 {\Gamma}_{\Lambda}^{(n)} ( \bd{k}_1 , \ldots , \bd{k}_n )
M_{\bd{k}_1 } \ldots M_{\bd{k}_n}  ,
 \hspace{7mm}
 \label{eq:GammaexpandIsing}
 \end{eqnarray}
where the Fourier coefficients of the magnetization field are defined by $ M_{\bd{k}} = \sum_i e^{ - i \bd{k} \cdot \bd{r}_i } M_i$.
Substituting the expansion (\ref{eq:GammaexpandIsing}) into the exact flow
equation (\ref{eq:Wetterich}) we obtain an infinite hierarchy of flow equations for the
$n$-point vertices.  For simplicity, we set  $h_0 =0$ 
and assume that there is no spontaneous magnetization. The flow equation for 
$\Gamma^{(2)}_{\Lambda} ( \bd{k} ) \equiv \Gamma^{(2)}_{\Lambda} ( - \bd{k} , \bd{k} )$ is then
 \begin{equation}
 \partial_{\Lambda} \Gamma^{(2)}_{\Lambda} ( \bd{k} ) = \frac{\beta}{2N}
 \sum_{\bd{q} } \dot{G}_{\Lambda} ( \bd{q} ) \Gamma^{(4)}_{\Lambda} ( - \bd{k} , \bd{k} , - \bd{q} , \bd{q} ),
 \label{eq:flow2}
 \end{equation}
where
 $
 \dot{G}_{\Lambda} ( \bd{k} )  =   - G^2_{\Lambda} ( \bd{k} ) \partial_{\Lambda} {R}_{\Lambda} ( \bd{k} ) $ is the so-called single-scale propagator 
and $ G_{\Lambda} ( \bd{k} )  =  [  \Gamma^{(2)}_{\Lambda} ( \bd{k} ) +  \beta {R}_{\Lambda} ( \bd{k} ) ]^{-1} $
is the regularized  propagator.
With our deformation scheme, the Fourier transform of the
regulator is
${R}_{\Lambda} ( \bd{k} ) =   ( 1 - \Lambda  )  V_{ \bd{k}}$, 
where $V_{\bd{k}} = 2 D J \gamma_{\bd{k}}$ is the Fourier transform of the exchange interaction and 
$\gamma_{\bd{k}} = D^{-1} \sum_{\mu =1}^D  \cos  (k_\mu a)$ is the
nearest-neighbor structure factor of a $D$-dimensional hypercubic lattice
with lattice spacing $a$.

To derive the initial condition at $\Lambda=0$, we note that
for vanishing exchange interaction 
the generating functional of the connected spin correlation functions
is
$ {\cal{G}}_0 [ h ] = \sum_i  B ( \beta  h_i )$,
where 
$ B ( y ) = \ln  [ \sinh ( ( S+1/2) y )/   \sinh ( y /2 ) ] $
is the primitive integral of the spin-$S$ Brillouin function $b ( y ) = d B ( y ) / dy$.
The initial value of the two-point vertex is therefore 
 $
 \Gamma_0^{(2)} ( \bd{k} ) = 1 / b^{\prime}   - \beta  V_{  \bd{k} },
 $
where $b^{\prime} = S ( S+1 )/3$ is the derivative of $b(y)$ at $y=0$.
The calculation of the initial functional $\Gamma_0 [ M ]$ requires the inversion of the  
Brillouin function which is not possible in closed form \cite{Kroeger15}.
However, we can iteratively calculate  the first few terms in the  vertex expansion.
For example, the initial value of the four-point vertex is
\begin{equation}
 \Gamma^{(4)}_{0} ( \bd{k}_1 , \bd{k}_2 , \bd{k}_3 , \bd{k}_4 ) 
 = - b^{\prime \prime \prime} /  ( b^{\prime} )^4 
 \equiv u_0 > 0,
 \end{equation}
where $b^{\prime \prime \prime } =[ 1- (2S+1)^4 ]/120 $ is the $3$rd derivative of $b(y)$ at $y=0$. In general, the initial values  $\Gamma^{(n)}_0$ of the higher-order vertices  can be expressed in terms of derivatives $b^{(m)}$ 
of the Brillouin function up to order $m \leq n-1$.

As a quantitative test 
of our deformation scheme, let us calculate the critical temperature $T_c$ of the spin-$S$ Ising model,
which can be identified with the temperature 
where $\Gamma^{(2)}_{\Lambda =1} ( 0 ) = 0$.  If we approximate the two-point vertex at vanishing momentum by its initial value $\Gamma^{(2)}_0 ( 0 ) = 1/b^{\prime} - \beta V_0$, we obtain 
the mean-field critical temperature $T_{c0}= 2 D J S ( S+1)/3$.
To go beyond mean-field theory, we need  a suitable truncation of the infinite 
hierarchy of FRG flow equations.
For simplicity, let us retain only the flowing two-point and four-point
vertices with their initial momentum dependence
and close the hierarchy by approximating  the
six-point vertex by its initial value $\Gamma^{(6)}_{0}$.
Our results for $T_c$ for $S=1/2$ and different dimensions $D$ are summarized
in Table~\ref{tab:Tcres}. 
\begin{table}\centering
	\ra{1.3}
	\begin{ruledtabular}
	\begin{tabular}{@{}cccccc@{}}
			{$D$} & \multicolumn{3}{c}{$T_c / T_{c0}$ for $S=1/2$} & \multicolumn{2}{c}{relative error in \%} \\ \cmidrule{2-4} \cmidrule{5-6}
			& {SFRG} & {$\mathcal{O} (D^{-1})$} & {benchmark} & {SFRG} & {$\mathcal{O} (D^{-1})$} \\ \midrule

			1 & 0 & 0 & 0 & 0 & 0 \\
			2 & 0 & 0.50& 0.57 & - & 12 \\
			3 & 0.744 & 0.79 & 0.752 & 1 & 5 \\
			4 & 0.839 & 0.85& 0.835 & 0.5 & 2 \\
			5 & 0.880 & 0.89& 0.878 & 0.3 & 1 \\
			6 & 0.904 & 0.908 & 0.903 & 0.2 & 0.6 \\
			7 & 0.920 & 0.923 & 0.919 & 0.1 & 0.4 \\
		\end{tabular}
		\caption{	
Comparison of our SFRG results for the critical temperature of the spin-$1/2$ 
Ising model to the accepted results (benchmark) \cite{Kramers41,Ferrenberg18,Lundow09,Butera12}.
The third column marked ${\cal{O}} ( D^{-1})$ is the prediction of our analytical
formula~(\ref{eq:Tc}).
		}
		\label{tab:Tcres}
	\end{ruledtabular}
\end{table}
Note that in $D =3$ our SFRG prediction for $T_c$ agrees 
with controlled
Monte Carlo results \cite{Ferrenberg18} with an accuracy of about $1 \%$, while
for $D > 3$ our SFRG result for $T_c$ is even more accurate.
For higher spins $S > 1/2$  
(not listed in Table~\ref{tab:Tcres}) we obtain $T_c$ with similar accuracy.

Obviously,  in two dimensions our truncated SFRG incorrectly predicts $T_c =0$, indicating
that in this case our simple truncation is not sufficient. 
Fortunately, we can formally use $1/D$ as a small parameter to develop a 
more systematic truncation strategy.
Using the fact that the Brillouin-zone average of
the $2n$-th power $\gamma^{2n}_{\bd{k}}$ of the structure factor is of the order $1/D^n$, we can iterate our hierarchy of flow equations to
generate a systematic expansion of
$\Gamma^{(2)}_{\Lambda =1} ( 0 )$ in powers of $1/D$.
By truncating this expansion at order $1/D$
and solving the resulting self-consistency equation for $T_c$ we obtain
 \begin{equation}
 \frac{ T_c}{T_{c0} }
 = \frac{1}{2} \left[ 1 + \sqrt{ 1 - \frac{u_0 (b^{\prime} )^2}{ D }} \right].
 \label{eq:Tc}
 \end{equation}
The values for  $T_c$ obtained
from this expression for $S=1/2$ are listed in the third column 
of Table~\ref{tab:Tcres}.
In two dimensions we now obtain a finite $T_c = T_{c0}/2$, but for $D \geq 3$
the $T_c$ obtained from our truncated SFRG 
turns out to be  more accurate than Eq.~(\ref{eq:Tc}).
We have also used our SFRG flow equations to generate the expansion of  
$\Gamma^{(2)}_{\Lambda =1} ( 0 )$ for arbitrary spin $S$
up to order $1/D^3$ \cite{Supplemental}; for  $D \geq 4$ the resulting estimate for $T_c$  
(not shown in Table~\ref{tab:Tcres}) significantly improves upon both the leading $1/D$ results and
the truncated SFRG results listed in Table~\ref{tab:Tcres}. Using a different truncation based on the derivative expansion, Machado and Dupuis obtained numerical results for $T_c$ in two and three dimensions with similar accuracy \cite{Machado10}.

{\it{Application to quantum spin systems.}} 
Let us now come back to the quantum Heisenberg Hamiltonian (\ref{eq:hamiltonian}).
The exact FRG flow of the generating functional of the 
irreducible spin vertices is then given by Eq.~(\ref{eq:Wetterich}).
By expanding both sides in powers of the components of the fluctuating magnetization 
$M^{\alpha}_i ( \tau )$,
we obtain the usual hierarchy of coupled FRG flow equations \cite{Kopietz10}.
However, a deformation scheme where initially
the exchange interaction is completely switched off cannot be used in this case, because then  the Legendre transform of the initial 
generating functional ${\cal{G}}_0 [ \bd{h} ]$ does not exist due to the lack of dynamics in the longitudinal fluctuations.
This problem has already been noticed by Ran\c con \cite{Rancon14}, who studied the $S=1/2$ XY model by expressing the spin operators in terms of hardcore bosons and then applying the lattice FRG developed in Refs.~[\onlinecite{Machado10,Rancon11A,*Rancon11B,*Rancon12A,*Rancon12B}]. For quantum Heisenberg models, there are several ways to avoid the problem of the non-existing Legendre transform for deformation schemes with initially decoupled sites.
One possibility is to choose the initial $J^{0}_{ij}$ such that for $\Lambda =0$ the system decouples into non-interacting dimers~\cite{Krieg18}, which  
is a convenient initial condition for spin systems with valence-bond ground 
states \cite{Read89,*Read90}. Alternatively, we can consider the flow of the amputated connected
spin correlation functions, which are generated by \cite{Kopietz10}
 \begin{equation}
 {\cal{F}}_{\Lambda} [ \bd{M} ] = {\cal{G}}_{\Lambda} \bigl[ - \sum_{j} J^{\Lambda}_{ij} \bd{M}_j \bigr]
- \frac{1}{2} \int_0^{\beta}  d \tau  
  \sum_{ij}  J^{\Lambda}_{ij}  \bd{M}_{i} \cdot \bd{M}_j .
  \label{eq:amputated_functional}
 \end{equation}
This functional satisfies the Polchinski equation \cite{Polchinski84},
 \begin{eqnarray}
\partial_{\Lambda}{{\cal{F}}}_{\Lambda} [ \bd{M} ]  
 &  = & 
\frac{1}{2} \int_0^{\beta} d \tau 
 \sum_{ij, \alpha} ( \partial_{\Lambda} {\mathbf{J}}^{-1}_{\Lambda} )_{ij} 
 \Biggl[\frac{ \delta^2 {\cal{F}}_{\Lambda} [ \bd{M} ] }{\delta M_i^{\alpha} ( \tau )   
\delta M_j^{\alpha} ( \tau ) }
 \nonumber
 \\
 &  & \hspace{-7mm} +
\frac{ \delta {\cal{F}}_{\Lambda} [ \bd{M} ] }{\delta M_i^{\alpha} ( \tau ) }
\frac{ \delta {\cal{F}}_{\Lambda} [ \bd{M} ] }{\delta M_j^{\alpha} ( \tau ) }
  \Biggr]
  + \frac{1}{2} {\rm Tr} \left[ \mathbf{J}_{\Lambda} \partial_\Lambda \mathbf{J}^{-1}_\Lambda \right],
 \hspace{7mm}
 \label{eq:flowGm}
 \end{eqnarray}
where $ \mathbf{J}^{-1}_{\Lambda} $ is the matrix inverse of $J_{ij}^{\Lambda}$.
The precise relation between our SFRG approach 
and the spin diagram technique developed by VLP \cite{Vaks68a,*Vaks68b} 
is established by 
the Legendre transform
${\Phi}_{\Lambda} [ \bd{h} ]$ of  ${\cal{F}}_{\Lambda} [ \bd{M} ]$,
which
satisfies a flow equation similar to Eq.~(\ref{eq:Wetterich})
and is well defined
even for vanishing exchange interaction \cite{Supplemental}.
In fact, in a scheme where $J^0_{ij} =0$,
the initial vertices generated by ${\Phi}_0 [ \bd{h} ]$
 can be identified with the generalized blocks 
introduced in Ref.~[\onlinecite{Izyumov88}].
These have a non-trivial frequency dependence \cite{Vaks68a,*Vaks68b,Izyumov88}  reflecting
the commutation relations between the components of $\bd{S}_i$ at a given site.
For finite $\Lambda$,
the functional
 ${\Phi}_{\Lambda} [ \bd{h} ]$
generates the part of the connected spin correlation functions which is irreducible with respect to
cutting a single interaction line. For the two-point function
this is precisely the irreducible self-energy calculated 
diagrammatically  by VLP \cite{Vaks68a,*Vaks68b}, see also Ref.~[\onlinecite{Izyumov88}].
In fact, by appropriately truncating the hierarchy of flow equations 
for the vertices generated by ${\Phi}_{\Lambda} [ \bd{h} ]$
we can recover, for example,  the expansion for the longitudinal spin-spin correlation 
function given by VLP \cite{Vaks68a,*Vaks68b,Supplemental}. However, in contrast to the perturbative approach of VLP, with a suitable truncation \cite{Berges02,Kopietz10} our SFRG  can also describe the critical regime. Furthermore, we can use our functional $\Phi_\Lambda [\bd{h}]$ to generalize our $1/D$ expansion to quantum spin systems. Considering the quantum Heisenberg model on a $D$-dimensional hypercubic lattice with nearest-neighbour interaction and retaining only the leading correction to $\Gamma^{(2)}_{\Lambda=1} (K)$, we find \cite{Supplemental}
\begin{equation}
\frac{ T_c}{T_{c0} }
= \frac{1}{2} \left[ 1 + \sqrt{ 1 - \frac{1}{D} \left[ \frac{5}{3} u_0 (b')^2 \pm \frac{1}{6 b'} \right] } \right].
\label{eq:Tc_quantum}
\end{equation}
The last term in the inner brackets is due to quantum effects and breaks the symmetry between a ferromagnetic (upper sign) and an antiferromagnetic (lower sign) exchange interaction, which is only restored in the classical limit $S \to \infty$. For an antiferromagnet with arbitary spin $S$, we find that the relative error of Eq.~\eqref{eq:Tc_quantum} is already below $10\%$ for $D=3$ \cite{Cuccoli01}, demonstrating that our SFRG is not restricted to ferromagnetic systems.

{\it{Summary and outlook.}}
The main result of this work is the insight that the generating functional of the
connected time-ordered spin correlation functions and the associated
generating functional of the irreducible vertices
 of an arbitrary quantum spin system
 satisfy exact flow equations, which are formally identical to the 
corresponding equations
of interacting bosons. 
The $SU(2)$ spin algebra is  
taken into account via a 
non-trivial initial condition involving vertices of arbitrary order.
At this point the full potential of our method has not been explored, but
our current results indicate that the SFRG
is a powerful analytical approach to quantum spin systems. In fact, we have recently shown how the one-loop scaling equations for the Kondo model can be obtained within the SFRG \cite{Tarasevych18}. Apart from offering an alternative to the unconventional renormalization of the $T$-matrix in Anderson's "poor man's scaling" approach \cite{Anderson70a}, the SFRG can also be extended to study the strong coupling regime or the electronic self-energy of the Kondo model.
Moreover, our SFRG  
can be easily generalized to any Hamiltonian which 
can be expressed in terms of local
operators satisfying a  non-trivial algebra such as
Hubbard X-operators~\cite{[{See, for example, }]Ovchinnikov04}.

We acknowledge discussions with N. Dupuis, A. Ran\c con, R. Thomale, O. Tsyplyatyev, and A. L. Chernychev, as well as the
hospitality of the Department of Physics and Astronomy of the University of California, 
Irvine, where part of this work was done.

\bibliographystyle{apsrev4-1}
\bibliography{srg}

\begin{titlepage}
	\vspace*{\stretch{1.0}}
	\begin{center}
		\large\textbf{Supplemental Material}\\
	\end{center}
	\vspace*{\stretch{2.0}}
\end{titlepage}

\setcounter{equation}{0}
\setcounter{figure}{0}
\setcounter{table}{0}
\setcounter{page}{1}
\renewcommand{\theHequation}{S\arabic{equation}}
\renewcommand{\theequation}{S\arabic{equation}}
\renewcommand{\theHfigure}{S\arabic{figure}}
\renewcommand{\thefigure}{S\arabic{figure}}
\renewcommand{\bibnumfmt}[1]{[S#1]}
\renewcommand{\citenumfont}[1]{S#1}

\section{Relation between spin FRG and the diagrammatic approach of Vaks, Larkin, and Pikin}

\subsection{General relations}

\begin{figure}
	\begin{center}
		\centering
		\includegraphics[width=\linewidth]{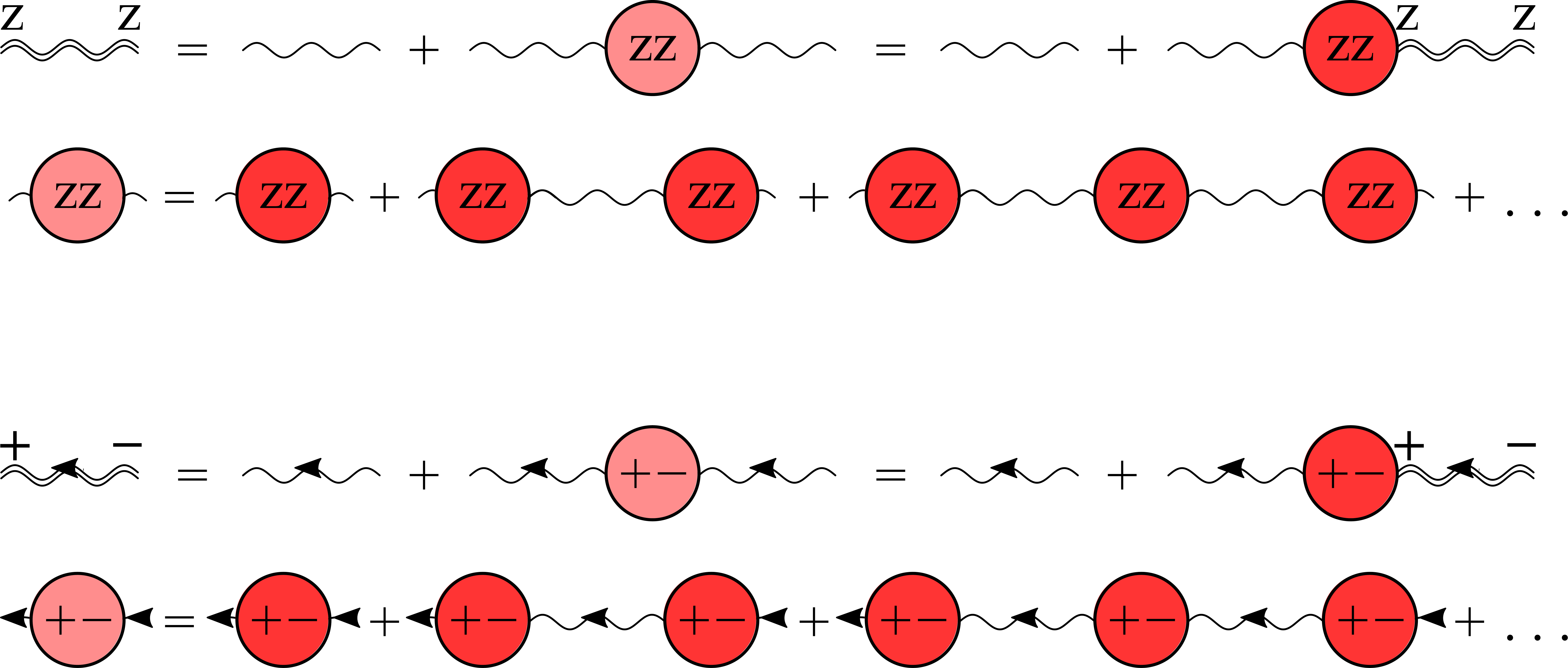}
	\end{center}
	\caption{
		Graphical representation of the relation between the longitudinal two-point functions $F^{zz}_\Lambda$, $G^{zz}_\Lambda$, and $\Pi^{zz}_\Lambda$ as given in Eqs.~\eqref{eq:sm_f_g_zz}, \eqref{eq:sm_f_pi_zz}, and \eqref{eq:propagator_pi_zz} (upper half) and the corresponding relation between the transverse two-point functions $F^{+-}_\Lambda$, $G^{+-}_\Lambda$, and $\Pi^{+-}_\Lambda$ as given in Eqs.~\eqref{eq:sm_f_g_+-}, \eqref{eq:sm_f_pi_+-}, and \eqref{eq:propagator_pi_+-} (lower half). Double wavy lines denote the amputated connected two-point functions $F^{\alpha \alpha'}_\Lambda$, while single wavy lines denote the deformed interaction $V_\Lambda$. The connected two-point functions $G^{\alpha \alpha'}_\Lambda$ are represented by light-colored circles, while the irreducible polarization functions $\Pi^{\alpha \alpha'}_\Lambda$ are represented by dark-colored circles.
	}
	\label{fig:sm_feynman_relations}
\end{figure}

In the first part of this Supplemental Material, we will give technical details on the exact relationship between our spin FRG and the diagrammatic approach to spin systems developed by Vaks, Larkin, and Pikin (VLP) \cite{aVaks68a,aVaks68b}. To adopt the notation of VLP, we set $J^\Lambda_{ij} = - V^\Lambda_{ij}$. The generating functional of the amputated connected spin correlation functions, $\mathcal{F}_\Lambda [\bd{M}]$, is then given by \cite{aKopietz10}
\begin{align}
e^{\mathcal{F}_\Lambda [\bd{M}]} = \text{Tr} \left[ e^{-\beta \mathcal{H}_0} \mathcal{T} e^{\frac{1}{2} \int_0^\beta d\tau \sum_{ij} V^\Lambda_{ij} (\bd{M}_i + \bd{S}_i) \cdot (\bd{M}_j + \bd{S}_j)} \right],
\label{eq:sm_def_F}
\end{align}
where ${\cal{H}}_0 = - h_0 \sum_i S^z_i$. Note that Eq.~\eqref{eq:sm_def_F} is equivalent to Eq.~\eqref{eq:amputated_functional} in the main text. To derive the flow equation of $\mathcal{F}_\Lambda [\bd{M}]$, it is helpful to first decouple the interaction term in Eq.~\eqref{eq:sm_def_F} via a three-component auxiliary field $\bd{\phi}_i (\tau)$,
\begin{widetext}
	\begin{align}
	e^{\mathcal{F}_\Lambda [\bd{M}]} = \frac{\int \mathcal{D} [\bd{\phi}] e^{-\frac{1}{2} \int_0^\beta d\tau \sum_{ij} [\mathbf{V}^{-1}_\Lambda]_{ij} \bd{\phi}_i (\tau) \cdot \bd{\phi}_j (\tau) + \int_0^\beta d\tau \sum_i \bd{M}_i (\tau) \cdot \bd{\phi}_i (\tau)} \text{Tr} \left[ e^{-\beta \mathcal{H}_0} \mathcal{T} e^{\int_0^\beta d\tau \sum_i \bd{\phi}_i (\tau) \cdot \bd{S}_i (\tau)} \right]}{\int \mathcal{D} [\bd{\phi}] e^{-\frac{1}{2} \int_0^\beta d\tau \sum_{ij} [\mathbf{V}^{-1}_\Lambda]_{ij} \bd{\phi}_i (\tau) \cdot \bd{\phi}_j (\tau)}},
	\label{eq:sm_F_HS}
	\end{align}
\end{widetext}
where $\mathbf{V}_\Lambda^{-1}$ is the matrix inverse of the matrix $[\mathbf{V}_\Lambda]_{ij} = V^\Lambda_{ij}$. By differentiating both sides of Eq.~\eqref{eq:sm_F_HS} with respect to $\Lambda$ we obtain the Polchinski equation \cite{aPolchinski84},
\begin{align}
\partial_{\Lambda}{{\cal{F}}}_{\Lambda} [ \bd{M} ]  
&= - \frac{1}{2} \int_0^{\beta} d \tau 
\sum_{ij, \alpha} ( \partial_{\Lambda} {\mathbf{V}}^{-1}_{\Lambda} )_{ij} 
\Biggl[\frac{ \delta^2 {\cal{F}}_{\Lambda} [ \bd{M} ] }{\delta M_i^{\alpha} ( \tau )   
	\delta M_j^{\alpha} ( \tau ) }
\nonumber
\\
&\hspace{-7mm} +
\frac{ \delta {\cal{F}}_{\Lambda} [ \bd{M} ] }{\delta M_i^{\alpha} ( \tau ) }
\frac{ \delta {\cal{F}}_{\Lambda} [ \bd{M} ] }{\delta M_j^{\alpha} ( \tau ) }
\Biggr]
+ \frac{1}{2} {\rm Tr} \left[ \mathbf{V}_{\Lambda} \partial_\Lambda \mathbf{V}^{-1}_\Lambda \right].
\hspace{7mm}
\label{eq:sm_flowGm}
\end{align}
To derive the FRG flow of the polarization functions con-
sidered by VLP, we introduce the subtracted Legendre
transform $\Phi_\Lambda [\bd{h}]$ of the functional $\mathcal{F}_\Lambda [\bd{M}]$,
\begin{align}
\Phi_\Lambda [\bd{h}] &= \int_0^\beta d\tau \sum_i \bd{M}_i (\tau) \cdot \bd{h}_i (\tau) - \mathcal{F}_\Lambda [\bd{M} [\bd{h}]]
\nonumber
\\
&- \frac{1}{2} \int_0^\beta d\tau \sum_{ij} \tilde{R}^\Lambda_{ij} \bd{h}_i (\tau) \cdot \bd{h}_j (\tau),
\label{eq:sm_phi_functional}
\end{align}
where the magnetization $\bd{M}_i (\tau)$ on the right-hand side should be considered as a functional of the source fields $\bd{h}_i (\tau)$ by inverting the relation
\begin{align}
h^\alpha_i (\tau) = \frac{\delta \mathcal{F}_\Lambda [\bd{M}]}{\delta M^\alpha_i (\tau)},
\end{align}
and the regulator $\tilde{R}^\Lambda_{ij}$ is defined in terms of the inverse exchange interaction,
\begin{align}
\tilde{R}^\Lambda_{ij} = [\mathbf{V}^{-1}_\Lambda]_{ij} - [\mathbf{V}^{-1}]_{ij},
\end{align}
where $\mathbf{V} = \mathbf{V}_{\Lambda=1}$ is the bare exchange interaction. Physically, the fields $\bd{h}_i (\tau)$ represent the exchange correction to the external magnetic field. The functional $\Phi_\Lambda [\bd{h}]$ satisfies the Wetterich equation \cite{aWetterich93}
\begin{align}
\partial_\Lambda \Phi_\Lambda [\bd{h}] &= \frac{1}{2} \text{Tr} \left\{ \left[ \left( \bd{\Phi}''_\Lambda [\bd{h}] + \mathbf{\tilde{R}}_\Lambda \right)^{-1} - \mathbf{V}_\Lambda \right] \partial_\Lambda \mathbf{\tilde{R}}_\Lambda \right\},
\end{align}
where $\bd{\Phi}''_\Lambda$ and $\mathbf{\tilde{R}}_\Lambda$ are matrices in all labels with matrix elements
\begin{align}
[\bd{\Phi}''_\Lambda [\bd{h}]]_{i\tau\alpha,j\tau'\alpha'} = \frac{\delta^2 \Phi_\Lambda [\bd{h}]}{\delta h^\alpha_i (\tau) \delta h^{\alpha'}_j (\tau')},
\nonumber
\\
[\mathbf{\tilde{R}}_\Lambda]_{i\tau\alpha,j\tau'\alpha'} = \tilde{R}^\Lambda_{ij} \delta_{\alpha \alpha'} \delta (\tau-\tau').
\end{align}
For finite external field or in the presence of a finite spontaneous magnetization, the functional $\Phi_\Lambda [\bd{h}]$ has a minimum at the scale-dependent uniform field configuration $\bd{h}_i = \bd{\bar{h}}_\Lambda$,
\begin{align}
\left. \frac{\delta \Phi_\Lambda [\bd{h}]}{\delta \bd{h}_i (\tau)} \right|_{\bd{h}=\bd{\bar{h}}_\Lambda} = 0,
\label{eq:sm_condition_minimum}
\end{align}
where $\bd{\bar{h}}_\Lambda$ is the cutoff-dependent exchange correction to the external magnetic field. It is then convenient to shift the fluctuating exchange field $\bd{h}_i (\tau) = \bd{\bar{h}}_\Lambda + \bd{\eta}_i (\tau)$ and consider the flow of the functional
\begin{align}
\tilde{\Phi}_\Lambda [\bd{\eta}] = \Phi_\Lambda [\bd{\bar{h}}_\Lambda + \bd{\eta}],
\end{align}
which is given by
\begin{align}
\partial_\Lambda \tilde{\Phi}_\Lambda [\bd{\eta}] &= \left. \partial_\Lambda \Phi_\Lambda [\bd{h}] \right|_{\bd{h} \to \bd{\bar{h}}_\Lambda + \bd{\eta}} + \int_0^\beta d\tau \sum_{i,\alpha} \frac{\delta \tilde{\Phi}_\Lambda [\bd{\eta}]}{\delta \eta^\alpha_i (\tau)} \partial_\Lambda \bar{h}^\alpha_\Lambda
\nonumber
\\
&= \frac{1}{2} \text{Tr} \left\{ \left[ \left( \bd{\tilde{\Phi}}''_\Lambda [\bd{\eta}] + \mathbf{\tilde{R}}_\Lambda \right)^{-1} - \mathbf{V}_\Lambda \right] \partial_\Lambda \mathbf{\tilde{R}}_\Lambda \right\}
\nonumber
\\
&+ \int_0^\beta d\tau \sum_{i,\alpha} \frac{\delta \tilde{\Phi}_\Lambda [\bd{\eta}]}{\delta \eta^\alpha_i (\tau)} \partial_\Lambda \bar{h}^\alpha_\Lambda.
\end{align}
To exhibit the relation of our spin FRG approach to the formalism developed by VLP \cite{aVaks68a,aVaks68b}, we first note that with Eq.~\eqref{eq:sm_def_F} we can express the regularized amputated connected two-point functions $F^{\alpha \alpha'}_\Lambda (K,K') \equiv \delta (K+K') F^{\alpha \alpha'}_\Lambda (K')$ through the regularized connected two-point functions $G^{\alpha \alpha'}_\Lambda (K,K') \equiv \delta (K+K') G^{\alpha \alpha'}_\Lambda (K)$ as follows,
\begin{subequations}
	\begin{align}
	F^{zz}_\Lambda (K) &= V_\Lambda (\bd{k}) + V_\Lambda (\bd{k}) G^{zz}_\Lambda (K) V_\Lambda (\bd{k}),
	\label{eq:sm_f_g_zz}
	\\
	F^{+-}_\Lambda (K) &= V_\Lambda (\bd{k}) + V_\Lambda (\bd{k}) G^{+-}_\Lambda (K) V_\Lambda (\bd{k}),
	\label{eq:sm_f_g_+-}
	\end{align}
\end{subequations}
where $K = (\bd{k},i\omega)$ is a collective label for momentum and Matsubara frequency, $\delta (K) = \beta N \delta_{\bd{k},0} \delta_{\omega,0}$, and $S^\pm = (S^x \pm i S^y)/\sqrt{2}$. Comparing these expressions with Eqs.~(16a) and (16b) of Ref.~[\onlinecite{aVaks68a}], we see that in the limit $\Lambda \to 1$ we can identify $F^{zz}_\Lambda (K)$ and $F^{+-}_\Lambda (K)$ with the effective interaction of VLP. This motivates the introduction of the polarization functions
\begin{subequations}
	\begin{align}
	\Pi^{zz}_\Lambda (K) &= V^{-1}_{\bd{k}} - \tilde{\Phi}^{zz}_\Lambda (K),
	\\
	\Pi^{+-}_\Lambda (K) &= V^{-1}_{\bd{k}} - \tilde{\Phi}^{+-}_\Lambda (K),
	\end{align}
\end{subequations}
where $\tilde{\Phi}^{\alpha \alpha'}_\Lambda (K,K') = \delta (K+K') \tilde{\Phi}^{\alpha \alpha'}_\Lambda (K)$. The regularized amputated connected two-point functions can then be written as
\begin{subequations}
	\begin{align}
	F^{zz}_\Lambda (K) &= \frac{1}{\tilde{\Phi}^{zz}_\Lambda (K) + \tilde{R}_\Lambda (\bd{k})} = \frac{V_\Lambda (\bd{k})}{1 - V_\Lambda (\bd{k}) \Pi^{zz}_\Lambda (K)},
	\label{eq:sm_f_pi_zz}
	\\
	F^{+-}_\Lambda (K) &= \frac{1}{\tilde{\Phi}^{+-}_\Lambda (K) + \tilde{R}_\Lambda (\bd{k})} = \frac{V_\Lambda (\bd{k})}{1 - V_\Lambda (\bd{k}) \Pi^{+-}_\Lambda (K)}.
	\label{eq:sm_f_pi_+-}
	\end{align}
\end{subequations}
It follows from Eqs.~\eqref{eq:sm_f_g_zz} and \eqref{eq:sm_f_g_+-} that the regularized connected two-point functions are given by
\begin{subequations}
	\begin{align}
	G^{zz}_\Lambda (K) &= \frac{\Pi^{zz}_\Lambda (K)}{1 - V_\Lambda (\bd{k}) \Pi^{zz}_\Lambda (K)},
	\label{eq:propagator_pi_zz}
	\\
	G^{+-}_\Lambda (K) &= \frac{\Pi^{+-}_\Lambda (K)}{1 - V_\Lambda (\bd{k}) \Pi^{+-}_\Lambda (K)}.
	\label{eq:propagator_pi_+-}
	\end{align}
\end{subequations}
In the limit $\Lambda \to 1$, the polarization functions $\Pi^{zz}_\Lambda (K)$ and $\Pi^{+-}_\Lambda (K)$ can thus be identified with the self-energies $\Sigma^{zz} (K)$ and $\Sigma^{+-} (K)$ introduced by VLP (cf. Eq.~(13) of Ref.~[\onlinecite{aVaks68a}]). A graphical representation of the relations between $F^{\alpha \alpha'}_\Lambda$, $G^{\alpha \alpha'}_\Lambda$, and $\Pi^{\alpha \alpha'}_\Lambda$ is shown in Fig.~\ref{fig:sm_feynman_relations}.

\subsection{Initial conditions}

Let us now discuss the initial condition of $\Phi_\Lambda [\bd{h}]$ in a deformation scheme where for $\Lambda = 0$ the exchange interaction is completely switched off $(V^{\Lambda=0}_{ij} = 0)$. For a uniform source field $\bd{\bar{h}} = \bar{h} \bd{e}_z$ along the direction of the external field, the initial functional $\Phi_0 [\bar{h}] = \Phi_{\Lambda=0} [\bar{h}]$ is
\begin{align}
\Phi_0 [\bar{h}] = N \left[ \frac{\beta \bar{h}^2}{2 V_0} - B[\beta (h_0 + \bar{h})] \right],
\end{align}
where $V_0 = V_{\bd{k}=0}$ and
\begin{align}
B(y) = \ln \left[ \frac{\sinh [(S + 1/2)y]}{\sinh (y/2)} \right]
\end{align}
is the primitive integral of the spin-$S$ Brillouin function $b(y)$, i.e.,
\begin{align}
B'(y) &= b(y)
\nonumber
\\
&= \left( S + \frac{1}{2} \right) \coth \left[ \left( S + \frac{1}{2} \right) y \right] - \frac{1}{2} \coth \left( \frac{y}{2} \right).
\end{align}
For $\Lambda = 0$, the condition~\eqref{eq:sm_condition_minimum} for the expectation value $\bar{h}_0$ of the exchange field therefore reduces to the self-consistency condition
\begin{align}
\bar{h}_0 = V_0 b [\beta (h_0 + \bar{h}_0)].
\end{align}
All correlation functions at $\Lambda = 0$ thus depend on the total magnetic field
\begin{align}
h = h_0 + \bar{h}_0.
\end{align}
With $\bar{h}_0 = V_0 M_0$ we obtain the usual mean-field self-consistency equation for the magnetization,
\begin{align}
M_0 = b [\beta (h_0 + V_0 M_0)],
\end{align}
corresponding to the zeroth-order result of VLP \cite{aVaks68a,aVaks68b}.

From Eqs.~\eqref{eq:propagator_pi_zz} and \eqref{eq:propagator_pi_+-} it is obvious that the polarization functions $\Pi^{zz}_\Lambda (K)$ and $\Pi^{+-}_\Lambda (K)$ are initially given by the connected two-point spin correlation functions,
\begin{subequations}
	\begin{align}
	\Pi^{zz}_0 (K) &= G^{zz}_0 (K),
	\\
	\Pi^{+-}_0 (K) &= G^{+-}_0 (K).
	\end{align}
\end{subequations}
Concerning higher-order coefficients of the functional $\tilde{\Phi}_\Lambda [\bd{\eta}]$, we find that at $\Lambda = 0$ they are simply related to the corresponding connected spin correlation functions of an isolated spin,
\begin{align}
\tilde{\Phi}^{\alpha_1 \ldots \alpha_n}_0 (K_1,\ldots,K_n) = - G^{\alpha_1 \ldots \alpha_n}_0 (K_1,\ldots,K_n).
\end{align}
In the diagrammatic approach of VLP, the connected spin correlation functions of an isolated spin are called blocks and are denoted by $\Gamma_n$; they can be calculated systematically using the generalized Wick theorem for spin operators \cite{aVaks68a,aVaks68b,aIzyumov88}.

\subsection{Leading correction to the free energy}

In their spin-diagrammatic approach to the three-dimensional Heisenberg model, VLP expand the free energy \cite{aVaks68a} as well as the transversal and the longitudinal self-energy \cite{aVaks68b} in powers of $1/r_0^3$, where $r_0$ is the range of the exchange interaction. Within our formulation of the spin FRG in terms of the functional $\tilde{\Phi}_\Lambda [\bd{\eta}]$, it is straightforward to recover the expansion of VLP by solving the flow equations iteratively and expanding in the number of momentum integrals. Let us first consider the exact flow equation of the regularized free energy in units of temperature,%$\beta F_\Lambda = \tilde{\Phi}_\Lambda [0]$,
\begin{align}
&\partial_\Lambda \tilde{\Phi}_\Lambda [0] = \frac{1}{2} \text{Tr} \left\{ \left[ \left( \bd{\tilde{\Phi}}''_\Lambda [0] + \mathbf{\tilde{R}}_\Lambda \right)^{-1} - \mathbf{V}_\Lambda \right] \partial_\Lambda \mathbf{\tilde{R}}_\Lambda \right\}
\nonumber
\\
&= - \frac{1}{2} \sum_{\bd{k},\omega} \left[ \frac{V_{\bd{k}} \Pi^{zz}_\Lambda (K)}{1 - \Lambda V_{\bd{k}} \Pi^{zz}_\Lambda (K)} + 2 \frac{V_{\bd{k}} \Pi^{+-}_\Lambda (K)}{1 - \Lambda V_{\bd{k}} \Pi^{+-}_\Lambda (K)} \right],
\end{align}
where we have used the deformation scheme $V_\Lambda (\bd{k}) = \Lambda V_{\bd{k}}$. To leading order we can replace the polarization functions $\Pi^{zz}_\Lambda (K)$ and $\Pi^{+-}_\Lambda (K)$ by their initial value, so that the leading correction to the free energy is given by
\begin{align}
\tilde{\Phi}_{\Lambda=1} [0] - \tilde{\Phi}_0 [0] &= \frac{1}{2} \sum_{\bd{k},\omega} \ln \left[ 1 - V_{\bd{k}} \Pi^{zz}_0 (K) \right]
\nonumber
\\
&+ \sum_{\bd{k},\omega} \ln \left[ 1 - V_{\bd{k}} \Pi^{+-}_0 (K) \right].
\end{align}
Since
\begin{align}
\Pi^{zz}_0 (K) &= \beta \delta_{\omega,0} b' (\beta h),
\quad
\Pi^{+-}_0 (K) &= \frac{b (\beta h)}{h - i \omega},
\end{align}
this expression is identical to Eq.~(17) of Ref.~[\onlinecite{aVaks68a}].

\subsection{Leading correction to the longitudinal polarization function}

\begin{figure*}
	\begin{center}
		\centering
		\includegraphics[width=\textwidth]{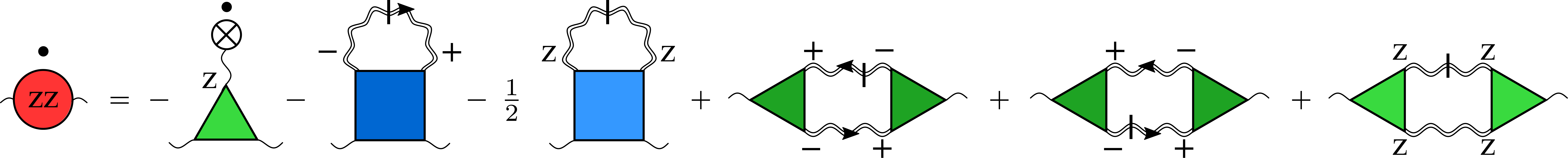}
	\end{center}
	\caption{
		Graphical representation of the flow equation for $\Pi^{zz}_\Lambda$ as given in Eq.~\eqref{eq:flow_exact_pi_zz}. Here the dot over the diagrams denotes the derivative $\partial_\Lambda$, the slashed double wavy lines represent the corresponding single-scale propagator, and the renormalized effective magnetic field $h_\Lambda$ is symbolized by a crossed circle. Except for the first term which is considered separately by VLP, the diagrams on the right-hand side correspond to Fig.~(3a) in Ref.~[\onlinecite{aVaks68b}] if we choose the deformation scheme $V_\Lambda (\bd{k}) = \Lambda V_{\bd{k}}$, replace the polarization functions as well as the higher-order vertices by their initial value, and integrate over $\Lambda$.
	}
	\label{fig:sm_flow_pi_zz}
\end{figure*}

In the same spirit, we can derive the $1/r_0^3$ expansion for higher-order coefficients of $\tilde{\Phi} [\bd{\eta}]$. Let us here consider $\Pi^{zz}_\Lambda (K)$ as a specific example, which satisfies the exact flow equation
\begin{widetext}
	\begin{align}
	\partial_\Lambda \Pi^{zz}_\Lambda (K) &= - \tilde{\Phi}^{zzz}_\Lambda (K,-K) \partial_\Lambda h_\Lambda - \int_{Q} \dot{F}^{+-}_\Lambda (Q) \tilde{\Phi}^{+-zz}_\Lambda (Q,-Q,K) - \frac{1}{2} \int_{Q} \dot{F}^{zz}_\Lambda (Q) \tilde{\Phi}^{zzzz}_\Lambda (Q,-Q,K)
	\nonumber
	\\
	&+ \int_Q \left[ \dot{F}^{+-}_\Lambda (Q) F^{+-}_\Lambda (Q+K) + F^{+-}_\Lambda (Q) \dot{F}^{+-}_\Lambda (Q+K) \right] \tilde{\Phi}^{+-z}_\Lambda (Q,-Q-K) \tilde{\Phi}^{+-z}_\Lambda (Q+K,-Q)
	\nonumber
	\\
	&+ \int_Q \dot{F}^{zz}_\Lambda (Q) F^{zz}_\Lambda (Q+K) \tilde{\Phi}^{zzz}_\Lambda (Q,-Q-K) \tilde{\Phi}^{zzz}_\Lambda (Q+K,-Q),
	\label{eq:flow_exact_pi_zz}
	\end{align}
\end{widetext}
where we have defined the single-scale propagators
\begin{subequations}
	\begin{align}
	\dot{F}^{zz}_\Lambda (K) &= - \left[ F^{zz}_\Lambda (K) \right]^2 \partial_\Lambda \tilde{R}_\Lambda (\bd{k})
	\nonumber
	\\
	&= \frac{\partial_\Lambda V_\Lambda (\bd{k})}{\left[ 1 - V_\Lambda (\bd{k}) \Pi^{zz}_\Lambda (K) \right]^2},
	\\
	\dot{F}^{+-}_\Lambda (K) &= - \left[ F^{+-}_\Lambda (K) \right]^2 \partial_\Lambda \tilde{R}_\Lambda (\bd{k})
	\nonumber
	\\
	&= \frac{\partial_\Lambda V_\Lambda (\bd{k})}{\left[ 1 - V_\Lambda (\bd{k}) \Pi^{+-}_\Lambda (K) \right]^2},
	\end{align}
\end{subequations}
and we have introduced the notation $\int_K = \frac{1}{\beta N} \sum_{\bd{k},\omega}$ as well as
\begin{align}
&\tilde{\Phi}^{\alpha_1 \ldots \alpha_n}_\Lambda (K_1,\ldots,K_n)
\nonumber
\\
= &\delta (K_1+\ldots+K_n) \tilde{\Phi}^{\alpha_1 \ldots \alpha_n}_\Lambda (K_1,\ldots,K_{n-1}).
\end{align}
A graphical representation of Eq.~\eqref{eq:flow_exact_pi_zz} is shown in Fig.~\ref{fig:sm_flow_pi_zz}.
We also need the flow equation of the renormalized effective magnetic field $h_\Lambda = h_0 + \bar{h}_\Lambda$,
\begin{align}
\tilde{\Phi}^{zz}_\Lambda (0) \partial_\Lambda h_\Lambda &= - \frac{1}{2} \int_Q \dot{F}^{zz}_\Lambda (Q) \tilde{\Phi}^{zzz}_\Lambda (Q,-Q)
\nonumber
\\
&- \int_Q \dot{F}^{+-}_\Lambda (Q) \tilde{\Phi}^{+-z}_\Lambda (Q,-Q).
\label{eq:sm_flow_magnetic_field}
\end{align}
To obtain the leading correction to $\Pi^{zz}_0$, we now approximate the polarization functions as well as the higher-order vertices in Eq.~\eqref{eq:sm_flow_magnetic_field} and on the right-hand side of Eq.~\eqref{eq:flow_exact_pi_zz} by their initial value. With the deformation scheme $V_\Lambda (\bd{k}) = \Lambda V_{\bd{k}}$, this allows us to perform the Matsubara sums as well as the integrals over $\Lambda$ analytically. We find that the first term on the right-hand side of Eq.~\eqref{eq:flow_exact_pi_zz} exactly reproduces the correction due to the renormalization of the magnetic field (see Eq.~(18) of Ref.~[\onlinecite{aVaks68a}]), while the remaining terms in Eq.~\eqref{eq:flow_exact_pi_zz} result in the correction given in Eq.~(36) of Ref.~[\onlinecite{aVaks68b}].

\section{\texorpdfstring{$1/D$}{1/D} expansion}

\subsection{Ising model}

In this section we expand on our discussion of the $1/D$ expansion in the main text. As we are interested in the critical temperature which is determined by $\Gamma^{(2)}_{\Lambda=1} = 0$, we need to solve the exact flow equation
\begin{align}
\partial_\Lambda \Gamma^{(2)}_\Lambda (\bd{k}) = \frac{\beta}{2 N} \sum_{\bd{q}} \frac{V_{\bd{q}} \Gamma^{(4)}_\Lambda (-\bd{k},\bd{k},-\bd{q},\bd{q})}{\left[ \Gamma^{(2)}_\Lambda (\bd{q}) + (1 - \Lambda) \beta V_{\bd{q}} \right]^2},
\end{align}
where we have assumed that $h_\Lambda = 0$ and we have again used the deformation scheme $V_\Lambda (\bd{k}) = \Lambda V_{\bd{k}}$. To leading order, we may approximate $\Gamma^{(4)}_\Lambda$ and $\Gamma^{(2)}_\Lambda$ by their initial values $u_0$ and $1/b' - \beta V_{\bd{k}}$, respectively. Expanding the denominator and using
\begin{align}
V_{\bd{q}} = V_0 \gamma_{\bd{q}} &= \frac{V_0}{D} \sum_{\mu=1}^D \cos (q_\mu a),
\\
\sum_{\bd{q}} \left( \gamma_{\bd{q}} \right)^{2n} &= \mathcal{O}(D^{-n}),
\end{align}
we arrive at
\begin{align}
\Gamma^{(2)}_{\Lambda=1} (\bd{k}) = \frac{1}{b'} \left[ 1 - \gamma_{\bd{k}} g + \frac{u_0 (b')^2}{4D} g^2 \right],
\end{align}
where the dimensionless parameter
\begin{align}
g = \beta b' V_0 = \frac{T_{c0}}{T} \text{sign} (V_0)
\end{align}
is defined in terms of the mean-field result for the critical temperature, $T_{c0} = b' |V_0|$. The condition that $\Gamma^{(2)}_{\Lambda=1}$ vanishes then results in Eq.~\eqref{eq:Tc} in the main text.
%in the critical temperature
%\begin{equation}
%\frac{ T_c}{T_{c0} }
%= \frac{1}{2} \left[ 1 + \sqrt{ 1 - \frac{u_0 (b^{\prime} )^2}{ D }} \right].
%\end{equation}
It is straightforward to go beyond this leading-order calculation. To next-to-leading order we also need to consider the flow of $\Gamma^{(4)}_\Lambda$ to first order in $1/D$ and insert the result in the flow equation of $\Gamma^{(2)}_\Lambda$. More generally, to $n$-th order in $1/D$ we have to take all irreducible vertices up to $\Gamma^{(2n)}_\Lambda$ into account. We have performed the expansion of $\Gamma^{(2)}_\Lambda$ up to third order, which for $V_0 > 0$ yields
\begin{align}
b' \Gamma^{(2)}_{\Lambda=1} (0) &= 1 - g + \frac{C_1 g^2}{D} - \frac{C_2 g^3}{D^2} + \frac{C_3 g^4}{D^2}
\nonumber
\\
&- \frac{C_4 g^4}{D^3} - \frac{C_5 g^5}{D^3} + \frac{C_6 g^6}{D^3}.
\label{eq:dim_exp_series_third}
\end{align}
The spin-dependent coefficients $C_i$ take on positive values of order unity and are explicitly given by
\begin{subequations}
	\begin{align}
	C_1 &= - \frac{b'''}{4 (b')^2},
	\\
	C_2 &= \frac{(b''')^2}{24 (b')^4},
	\\
	C_3 &= - \frac{b^{(5)} + 12 b' b'''}{32 (b')^3},
	\\
	C_4 &= \frac{b''' b^{(5)} - 36 (b')^3 b'''}{192 (b')^5},
	\\
	C_5 &= \frac{b' b''' b^{(5)} - (b''')^3 + 9 (b')^2 (b''')^2}{48 (b')^6},
	\\
	C_6 &= - \frac{1}{384 (b')^5} \bigg[ b' b^{(7)} + 3 b''' b^{(5)} + 36 (b')^2 b^{(5)}
	\nonumber
	\\
	&\hspace{22mm} + 80 b' (b''')^2 + 360 (b')^3 b''' \bigg],
	\end{align}
\end{subequations}
where $b^{(n)}$ is the $n$-th derivative of the Brillouin function $b(y)$ at $y=0$. For the special case $S=1/2$ this yields
\begin{align}
\frac{1}{4} \Gamma^{(2)}_{\Lambda=1} (0) &= 1 - g + \frac{g^2}{2 D} - \frac{g^3}{6 D^2} + \frac{g^4}{4 D^2}
\nonumber
\\
&- \frac{5 g^4}{24 D^3} - \frac{g^5}{4 D^3} + \frac{g^6}{2 D^3}.
\end{align}

\subsection{Quantum Heisenberg model}

As noted in the main text, for the quantum Heisenberg model the generating functional of the irreducible vertices, $\Gamma_\Lambda [\bd{M}]$, does not exist for vanishing exchange interaction due to the absence of longitudinal spin dynamics. We therefore cannot directly carry over our approach from the Ising model. However, the Legendre transform of the generating functional of the amputated connected spin correlation functions, $\Phi_\Lambda [\bd{h}]$, is well defined for $V^{\Lambda=0}_{ij}=0$. Again assuming a $D$-dimensional hypercubic lattice with nearest-neighbor interaction, we can expand the polarization function $\Pi_\Lambda$ in powers of $1/D$ in the same way as we did for the irreducible two-point vertex in the Ising model. Only at the end we invert $\Pi_{\Lambda=1}$ to generate the $1/D$ expansion of $\Gamma^{(2)}_{\Lambda=1}$. Compared to the Ising model, the calculations are now more complicated due to the additional frequency dependence as well as due to the larger number of finite connected spin correlators; however, this does not pose any conceptual difficulties. An advantage of our approach is that the diagrammatic expansion is given by the familiar expansion of the Wetterich equation. Assuming $h_\Lambda = 0$ so that we can define $\Gamma^{\alpha \alpha'}_\Lambda (K,K') \equiv \delta (K+K') \delta_{\alpha,\alpha'} \Gamma^{(2)}_\Lambda (K')$, we find to leading order in $1/D$
\begin{align}
\Gamma^{(2)}_{\Lambda=1} (K) + V_{\bd{k}} &= \frac{\delta_{\omega,0}}{\beta b'} \left[ 1 + \frac{g^2}{24 D} \left( 10 u_0 (b')^2 + \frac{\gamma_{\bd{k}}}{b'} \right) \right]
\nonumber
\\
&+ (1 - \delta_{\omega,0}) \frac{\beta \omega^2 D}{(1 - \gamma_{\bd{k}}) g^2},
\end{align}
where the momentum dependence of the corrections arises from the non-commutativity of the spin operators via the finite three-point vertex $\left< S^x S^y S^z \right>$; this $\bd{k}$ dependence also results in an asymmetry in the critical temperature with respect to the sign of the exchange interaction, as noted in the main text after Eq.~\eqref{eq:Tc_quantum}. Extending our expansion to next-to-leading order, we find for the static part of $\Gamma^{(2)}_{\Lambda=1}$ in the quantum limit $S=1/2$
\begin{align}
\frac{\beta}{4} \Gamma^{(2)}_{\Lambda=1} (\bd{k},0) &= 1 - \gamma_{\bd{k}} g + \frac{(5 + \gamma_{\bd{k}}) g^2}{6 D}
\nonumber
\\
& - \frac{(1 + \gamma_{\bd{k}}) g^3}{3 D^2}+ \frac{(1 - \gamma_{\bd{k}})^2 g^4}{36 D^2}.
\end{align}
Setting $D=3$ and $\gamma_{\bd{k}} = \text{sign} (V_0)$, we find that this result is consistent with the high-temperature series for the (staggered) susceptibility as given in Ref.~[\onlinecite{aOitmaa04}].

\end{document}